\documentclass[prl,twocolumn,superscriptaddress,showpacs,floatfix]{revtex4-1}
\usepackage[ulem=normalem]{changes}
\usepackage{graphicx,amssymb}
\usepackage[fleqn]{amsmath}
\usepackage{amsfonts}
\usepackage{dcolumn}
\usepackage[lmargin=2.1cm, rmargin=2.1cm,tmargin=3.4cm,bmargin=4.1cm]{geometry}
\usepackage{bm}
\usepackage{braket}
\usepackage{multirow} 
\usepackage{MnSymbol}

\usepackage{xcolor}
\RequirePackage[dvips]{hyperref}
\hypersetup
{
		bookmarks=true,%
		unicode=false,%
		pdftoolbar=true,
		pdfmenubar=true,
		pdffitwindow=false,
		pdfstartview=FitH,
		pdftitle={PRL},
		pdfauthor={J. Oko{\l}owicz},
		pdfsubject={Nuclear Physics},
		pdfcreator={J. Oko{\l}owicz},
		pdfproducer={J. Oko{\l}owicz},
		pdfkeywords={CSM},		
            bookmarksnumbered,%
		colorlinks=true,%
		linkcolor=blue,%
		citecolor=red,%
		filecolor=green,%
		urlcolor=blue,
		breaklinks, 
		plainpages=false%
}

\begin{document}

\preprint{ }

\title{Consistent analysis of one-nucleon spectroscopic factors involving weakly- and strongly-bound nucleons }

\author{J. Oko{\l}owicz}
\affiliation{Institute of Nuclear Physics, Polish Academy of Sciences, Radzikowskiego 152, PL-31342 Krak{\'o}w, Poland}

\author{Y.H. Lam}
\affiliation{Key Laboratory of High Precision Nuclear Spectroscopy,
Institute of Modern Physics, Chinese Academy of Sciences, Lanzhou 730000, China}

\author{M. P{\l}oszajczak}
\affiliation{Grand Acc\'el\'erateur National d'Ions Lourds (GANIL), CEA/DSM - CNRS/IN2P3,
BP 55027, F-14076 Caen Cedex, France}
 
\author{A.O. Macchiavelli}
\affiliation{Nuclear Science Division, Lawrence Berkeley National Laboratory, Berkeley, California 94720, USA}

\author{N.A. Smirnova}
\affiliation{CENBG (UMR 5797 - Universit{\'e} Bordeaux 1 - CNRS/IN2P3),
Chemin du Solarium, Le Haut Vigneau, BP 120, 33175 Gradignan Cedex, France}

\date{\today}

\begin{abstract}
	{ There is a considerable interest in understanding the dependence of one-nucleon removal cross sections on the asymmetry of the neutron $S_n$ and proton $S_p$ separation energies, following a large amount of experimental data and theoretical analyses in a framework of sudden and eikonal approximations of the reaction dynamics. These theoretical calculations involve both the single-particle cross section and the shell-model description of the projectile initial state and final states of the reaction residues. The configuration mixing in shell-model description of nuclear states depends on the proximity of one-nucleon decay threshold but does it depend sensitively on $S_n - S_p$? To answer this question, we use the shell model embedded in the continuum to investigate the dependence of one-nucleon spectroscopic factors on the asymmetry of $S_n$ and $S_p$ 
for mirror nuclei $^{24}$Si, $^{24}$Ne and $^{28}$S, $^{28}$Mg and for a series of neon isotopes ($20 \leq A \leq 28$).
			}
\end{abstract}


\maketitle

\emph{Introduction}.---
\label{sec0}
Single-nucleon removal reactions at intermediate energies provide a basic tool to produce exotic nuclei with large cross-sections. The interpretation of these reactions as effective direct reactions follows from the theoretical modelling which uses an approximate description of the reaction dynamics (sudden and eikonal approximations) \cite{theory_tost} and a standard shell model description of the structure of the initial state of the projectile, the final states of the $(A-1)$-nucleon reaction residues, and the relevant overlap functions. 
In a series of papers \cite{Gade1,Gade2}, it was found that the ratio $R_{\sigma}=\sigma_{\rm exp}/\sigma_{\rm th}$ of the experimental and theoretical inclusive one-nucleon removal cross section for a large number of projectiles shows a striking dependence on the asymmetry of the neutron and proton separation energies. Is this dependence telling us something important about the correlations in the projectile initial state and/or final states of reaction residues, or could it be an artefact of the theoretical modelling used? 

A possible drawback of the theoretical modelling of the one-nucleon removal reactions is that the description of reaction dynamics and shell model ingredients in this description are not consistently related one to another. Therefore, there is no {\it a priori} certainty that these reactions   can be considered as effective direct reactions. A comprehensive theoretical description in a unified framework of the continuum shell model (CSM) \cite{mahaux69_b16,barz77_82,bennaceur00_44} or the coupled-channel Gamow shell model \cite{Jag14} is much too complicated to be considered as a realistic proposition in the near future. On the other hand, single-nucleon removal reactions as experimental tools to study exotic nuclei are too important to abandon further discussion of $R_{\sigma}(\Delta S)$-dependence found, where $\Delta S$ equals $S_n-S_p$ for one-neutron removal and $S_p-S_n$ for one-proton removal reactions.  Recently, spectroscopic factors for proton knockout in closed-shell $^{14,16,22,24,28}$O were calculated using the coupled cluster formalism  \cite{Hagen} and found to depend  strongly on $\Delta S$, in line with the observations in \cite{Gade1, Gade2}.  In contrast,  the experimental results in Ref. \cite{Flavigny} using transfer reactions in Oxygen isotopes show, at best, a weak dependence on $\Delta S$.  Moreover, the study in Ref. \cite{Obertelli} points to some limitations of the eikonal approximation.
Here we would like to address the question of whether the deduced ratio of the experimental and theoretical one-nucleon removal cross-sections is related in any way to the $\Delta S$-dependence of the CSM spectroscopic factors? Hence, is this ratio probing the configuration mixing in SM states involved in the single-nucleon removal reactions at intermediate energies or should it be considered as an empirical normalization factor of the theoretical cross-section when determining spectroscopic factors of exotic nuclei? Of course, one should keep in mind that the spectroscopic factors are not observables {\it per se} and as such are not invariant under the unitary transformation of the Hamiltonian. However, in a given model, the spectroscopic factors are important theoretical quantities and an investigation of the $\Delta S$-dependence of the spectroscopic factors in a consistent theoretical framework provided by the CSM may shed light on our understanding of the one-nucleon removal reactions at intermediate energies. 

The $\Delta S$-dependence of spectroscopic factors is examined in $^{24}$Si and $^{28}$S, and their mirror partners $^{24}$Ne and $^{28}$Mg using the shell model embedded in the continuum (SMEC) \cite{bennaceur00_44} which is a modern version of the CSM. The ratio $R_{\sigma}$, which was reported for $^{24}$Si and $^{28}$S both for neutron and proton removal reactions \cite{Gade2}, will serve as a reference in this analysis. We investigate also the $\Delta S$-dependence of the ratio of SMEC and SM spectroscopic factors $R_{SF}=S_{\rm SMEC}/S_{\rm SM}$ for a chain of neon isotopes at the experimental separation energies $S_n$ and $S_p$.

\emph{The model}.---
In the present studies, the scattering environment is provided by one-nucleon decay channels. The Hilbert space is divided into two orthogonal subspaces ${\cal Q}_{0}$ and ${\cal Q}_{1}$ containing 0 and 1 particle in the scattering continuum, respectively. An open quantum system description of ${\cal Q}_0$  space includes couplings to the environment of decay channels through the energy-dependent effective Hamiltonian \cite{bennaceur00_44,Oko1}:
\begin{equation}
{\cal H}(E)=H_{{\cal Q}_0{\cal Q}_0}+W_{{\cal Q}_0{\cal Q}_0}(E),
\label{eq21}
\end{equation}
where $H_{{\cal Q}_0{\cal Q}_0}$ denotes the standard SM Hamiltonian describing the internal dynamics in the closed quantum system approximation, and $W_{{\cal Q}_0{\cal Q}_0}(E)$:
\begin{equation}
W_{{\cal Q}_0{\cal Q}_0}(E)=H_{{\cal Q}_0{\cal Q}_1}G_{{\cal Q}_1}^{(+)}(E)H_{{\cal Q}_1{\cal Q}_0},
\label{eqop4}
\end{equation}
is the energy-dependent continuum coupling term, where $G_{{\cal Q}_1}^{(+)}(E)$ is the one-nucleon Green's function and ${H}_{{Q}_0,{Q}_1}$, ${H}_{{Q}_1{Q}_0}$ are the coupling terms between orthogonal subspaces ${\cal Q}_{0}$ and ${\cal Q}_{1}$ \cite{Oko1}.
 $E$ in the above equations  stands for a scattering energy. The energy scale is settled by the lowest one-nucleon emission threshold. The channel state in nucleus $A$ is defined by the coupling of one nucleon (proton or neutron) in the continuum to nucleus  $(A-1)$ in a given SM state. In the SMEC calculation, we include 18 lowest decay channels in the nucleus $A$: 9 for protons and 9 for neutrons. The coupling to these channels gives rise to the mixing of all SM states of the same total angular momentum and parity in the nucleus $A$ and, hence changes the ground state (g.s.) spectroscopic factor with respect to its SM value.
 
The SMEC solutions are found by solving the eigenproblem for the Hamiltonian Eq.(\ref{eq21}) in the biorthogonal basis.
The SMEC eigenvectors $\Psi_{\alpha}$ are related to the eigenstates $\Phi_j$ of the SM Hamiltonian $H_{{\cal Q}_0{\cal Q}_0}$ by a linear orthogonal transformation: $\Psi_{\alpha}=\sum_jb_{\alpha j}\Phi_j$. The expectation value of any operator ${\hat O}$ can be calculated as: $\langle{\hat O}\rangle=\langle{\Psi}_{{\bar \alpha}}|{\hat O}|{\Psi}_{\alpha}\rangle$. In case of the spectroscopic factor one has: ${\hat O}=a^{\dagger}|t\rangle\langle t|a$, where $|t\rangle$ is the target state of the $(A-1)$-system. $a$ and $a^{\dagger}$ are annihilation and creation operators of a nucleon in a given single-particle (s.p.) state.

\emph{The Hamiltonian}.---
For the isospin-symmetric part of the SM Hamiltonian $H_{{\cal Q}_0{\cal Q}_0}$ in the $sd$ shell model space, 
we take the USD interaction \cite{Brown}. 
The charge-dependent terms in the Hamiltonian comprise the two-body Coulomb interaction, acting between valence protons, and 
isovector single-particle energies~\cite{Ormand} which account for the Coulomb effects in the core. 
Both these terms are scaled proportionally to $\sqrt{\hbar\omega_A}$~\cite{Ormand}, 
with $\hbar\omega_A$ being parameterized as 
\begin{equation}
\hbar \omega_A = 45 \, A^{-1/3} - 25 \, A^{-2/3}  ~ \textnormal{MeV} \, .
\end{equation}

The terms ${H}_{{Q}_0,{Q}_1}$, ${H}_{{Q}_1{Q}_0}$ in the continuum-coupling term Eq.(\ref{eqop4}) are generated using the Wigner-Bartlett (WB) interaction:
\begin{equation}
V_{12}=V_0\left[\alpha+ \beta P_{12}^{\sigma}\right]\delta\left({\bf r_1}-{\bf r_2}\right),
\label{WB}
\end{equation}
where $\alpha + \beta = 1$ and $P_{12}^{\sigma}$ is the spin exchange operator. 
Although the product $V_0(\alpha - \beta)$= 414 MeV$\cdot$fm$^3$ is kept constant in all calculations,
the magnitude of the continuum coupling varies depending on specific values of $V_0$, $\alpha $, 
the structure of the target wave function, and the separation energies $S_p$, $S_n$.

The radial s.p. wave functions in ${\cal Q}_0$ and the scattering wave functions in ${\cal Q}_1$ are generated using a Woods-Saxon (WS) potential which includes spin-orbit and Coulomb parts. The radius and diffuseness of the WS potential are $R_0=1.27 A^{1/3}$ fm and $a=0.67$ fm respectively. The spin-orbit potential is $V_{\rm SO}=6.1$ MeV,  and the Coulomb part is calculated for a uniformly charged sphere with radius $R_0$.  The depth of the central part for protons (neutrons) is adjusted to yield the energy of the s.p. state involved in the lowest one-proton (one-neutron) decay channel equal to the one-proton (one-neutron) separation energy in the g.s. of the nucleus $A$. The continuum-coupling term Eq.(\ref{eqop4}) breaks the isospin conservation due to different radial wave functions for protons and neutrons, and different separation energies $S_p$, $S_n$.

\begin{figure}[t!]
	\includegraphics[width=0.80\linewidth]{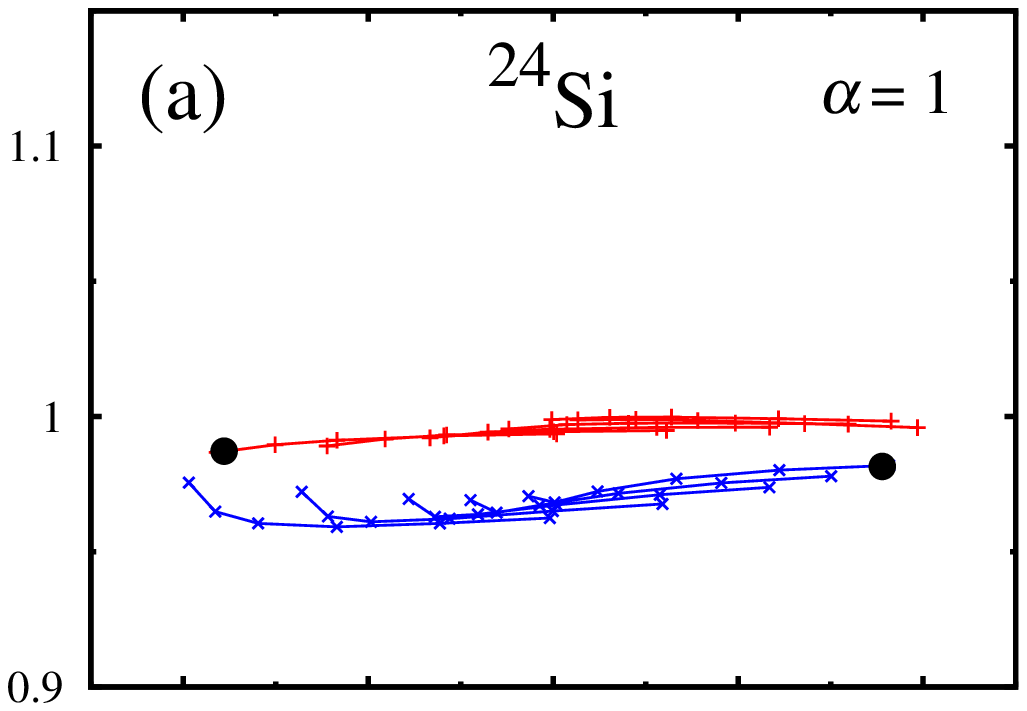}
\vskip -1.11truecm
 	\includegraphics[width=0.80\linewidth]{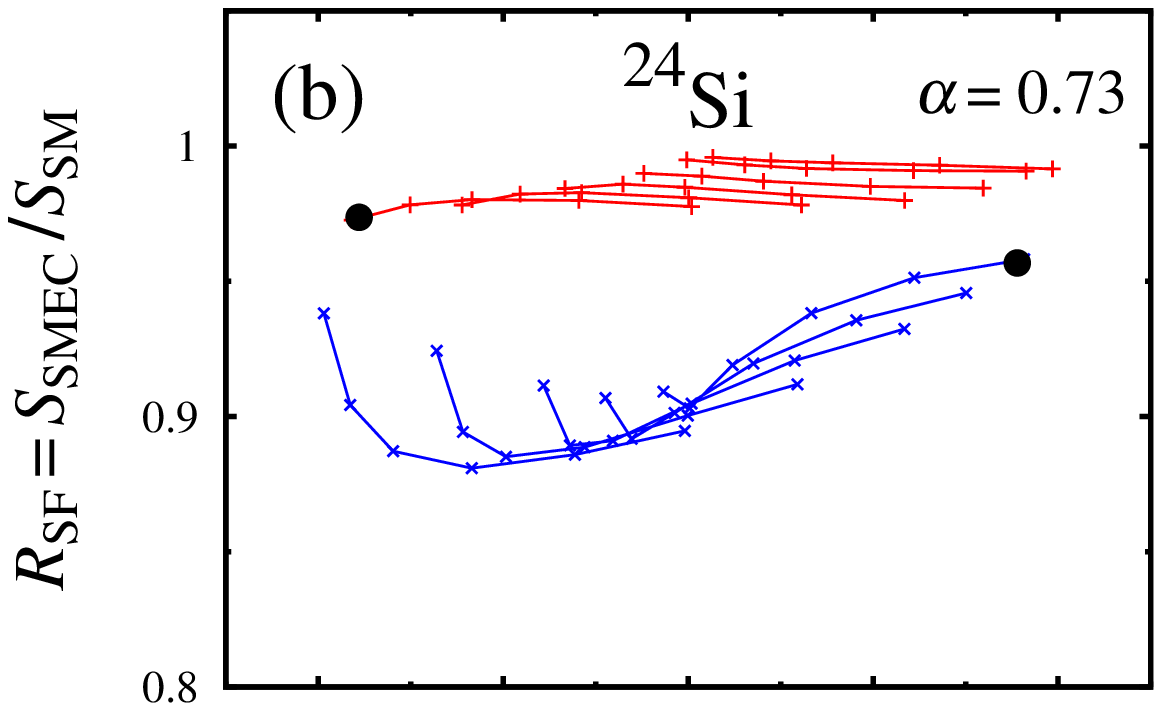}
\vskip -1.11truecm
 	\includegraphics[width=0.80\linewidth]{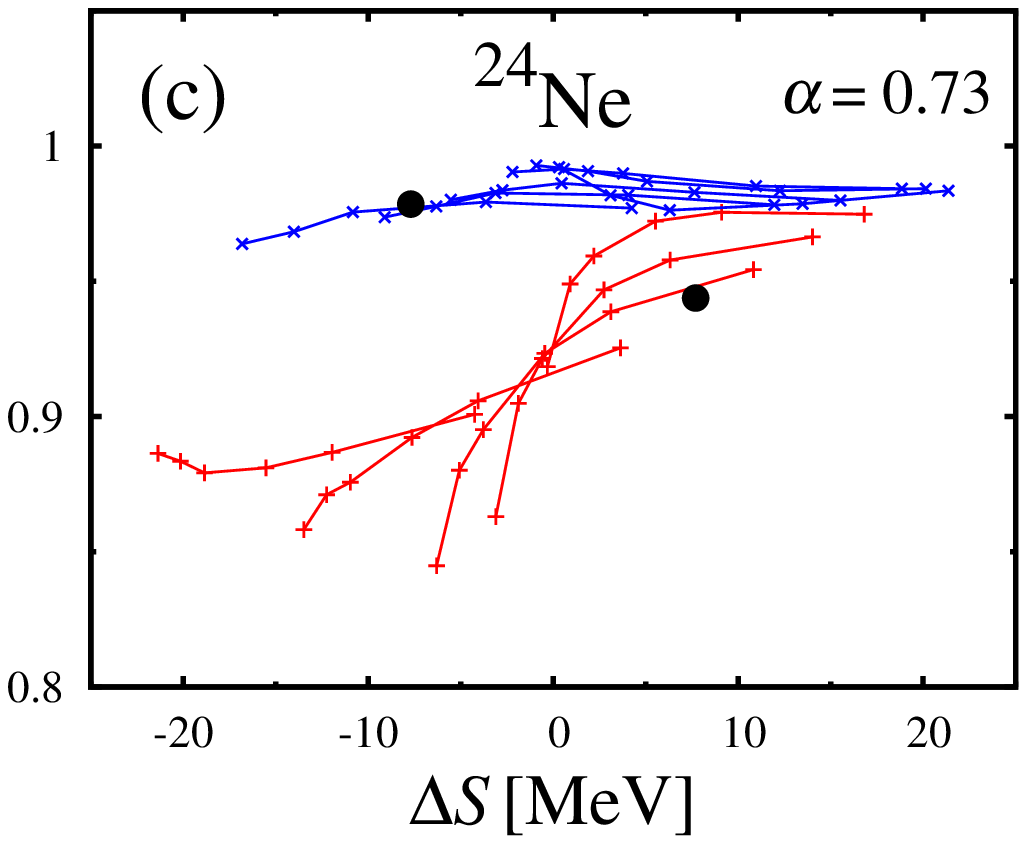}
\caption{(Color online) The ratio of SMEC and SM $d_{5/2}$ spectroscopic factors in mirror nuclei $^{24}$Si and $^{24}$Ne is plotted as a function of the asymmetry $\Delta S$ of the neutron and proton separation energies. Proton (neutron) spectroscopic factors are shown by solid (dashed) lines. $\Delta S$ equals $S_p-S_n$ ($S_n-S_p$) for the proton (neutron) spectroscopic factors. 
Circles denote the ratios $R_{SF}$ at the experimental value of the asymmetry parameter $\Delta S$. 
(a) $R_{SF}(\Delta S)$ for the g.s. of $^{24}$Si for $\alpha =1$; 
(b) $R_{SF}(\Delta S)$ for the g.s. of $^{24}$Si for $\alpha=0.73$; 
(c) $R_{SF}(\Delta S)$ for the g.s. of $^{24}$Ne for 
$\alpha =0.73$. For more details, see the description in the text.}
	\label{fig_1}
\end{figure}

\begin{figure}[t!]
        \includegraphics[width=0.8\linewidth]{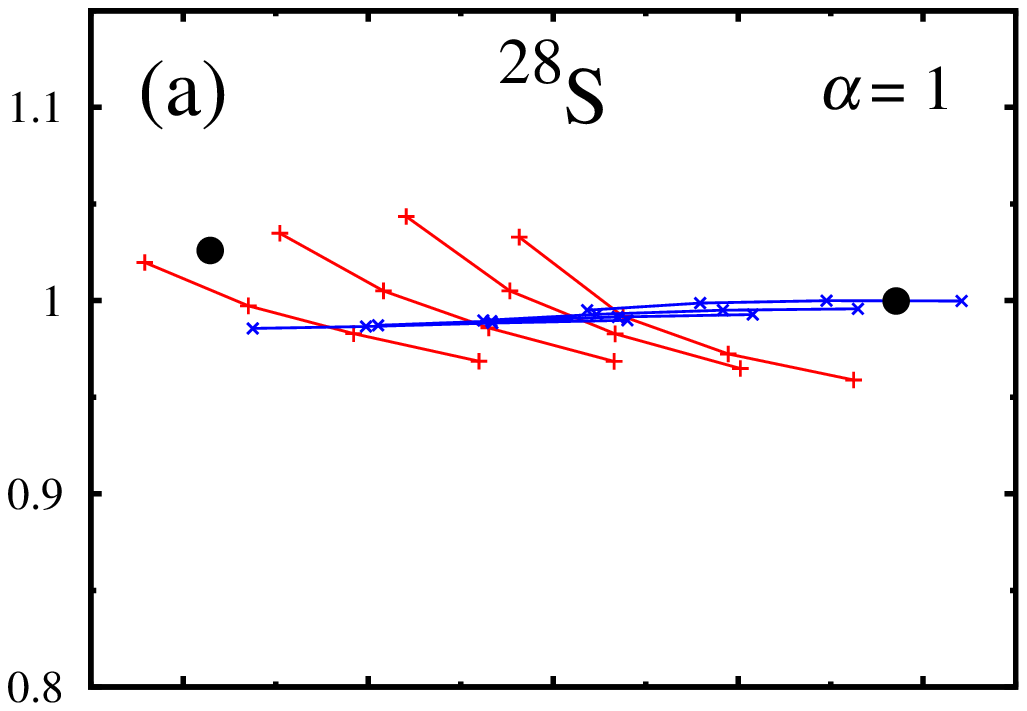}
\vskip -1.11truecm
         \includegraphics[width=0.8\linewidth]{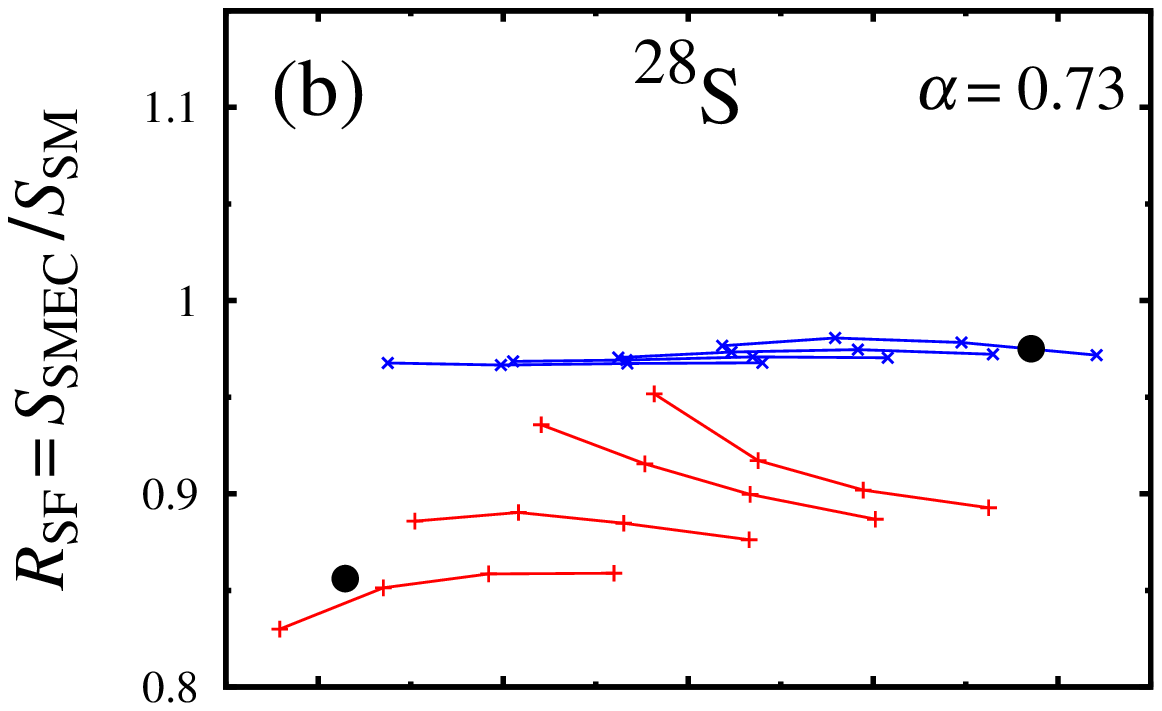}
\vskip -1.11truecm
         \includegraphics[width=0.8\linewidth]{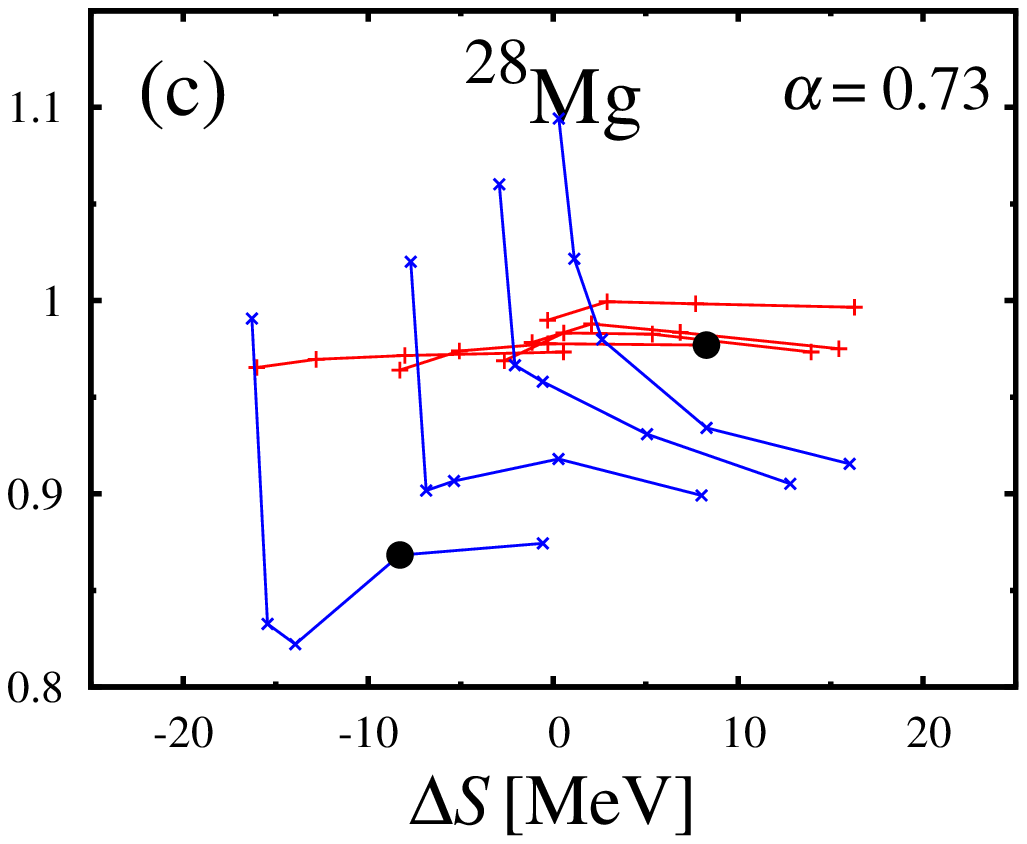}
         \caption{(Color online) The ratio of SMEC and SM proton (neutron) $s_{1/2}$ spectroscopic factors and neutron (proton) $d_{5/2}$ spectroscopic factors in mirror nuclei $^{28}$S ($^{28}$Mg) is plotted as a function of the asymmetry $\Delta S$ of the neutron and proton separation energies.  (a) $R_{SF}(\Delta S)$ for the g.s. of $^{28}$S for $\alpha =1$; 
(b) $R_{SF}(\Delta S)$ for the g.s. of $^{28}$Si for $\alpha =0.73$; 
(c) $R_{SF}(\Delta S)$ for the g.s. of $^{28}$Mg for $\alpha =0.73$. 
For more information, see the caption of Fig. \ref{fig_1} and the description in the text.}
	\label{fig_2}
\end{figure}

{\emph{The results}.---}
We shall discuss now the dependence of the spectroscopic factors on the asymmetry of neutron and proton separation energies and 
on the strength of the spin-exchange term which influences the strength of the $T=0$ proton-neutron continuum coupling. 
Fig. \ref{fig_1} shows the $\Delta S$ dependence of the ratio of $d_{5/2}$ spectroscopic factors in SMEC and SM for the g.s. of mirror nuclei: $^{24}$Si and $^{24}$Ne. The curves $R_{SF}(\Delta S)$ for proton (neutron) spectroscopic factors are shown with the solid (dashed) lines as a function of $S_p-S_n$ ($S_n-S_p$) on a $(S_p,S_n)$-lattice. The values of $R_{SF}$ at the experimental separation energies $S_p$ and $S_n$ are denoted by open circles. Along each curve for the proton (neutron) spectroscopic factor, $S_p$ ($S_n$) changes and $S_n$ ($S_p$) remains constant. 
 
Quantitative effects of the continuum coupling on spectroscopic factors depend on the distribution of the spectroscopic strength in the considered SM states \cite{OMNP12}. For $\alpha= 1$, the rearrangement of spectroscopic factors is small if the spectroscopic strength is concentrated in a single SM state. This is the case in both $^{24}$Si and $^{24}$Ne. For example, 92.7$\%$ (91.9$\%$) of the proton (neutron) $d_{5/2}$ SM spectroscopic strength in $^{24}$Si is in the g.s. and hence, the ratio $R_{SF}$ both for proton and neutron g.s. spectroscopic factors is almost independent of $\Delta S$ (see Fig. 1a). The slight breaking of the isospin symmetry by both the continuum coupling and the Coulomb term in the SM interaction leads in this case to a weak dependence of $R_{SF}$ for the neutron spectroscopic factor in the limit of small $S_n$.

This is not true anymore if the continuum-coupling interaction includes 
the spin-exchange term, which modifies the $T=0$ proton-neutron continuum coupling component. In this case, the additional correlations induced via the coupling to decay channels may modify the ratio of spectroscopic factors. Figs. 1b and 1c show a rearrangement of the $d_{5/2}$ neutron (proton) spectroscopic strength in the g.s. of $^{24}$Si ($^{24}$Ne) for $\alpha =0.73$. 
The $\Delta S$-variation of the ratio $R_{SF}$ in this case remains however relatively small and does not exceed 10$\%$. 
One should notice also a significant breaking of mirror symmetry by comparing $R_{SF}(\Delta S)$ curves for neutron spectroscopic factors at small $S_n$ in $^{24}$Si and proton spectroscopic factors at small $S_p$ in $^{24}$Ne.

The ratio $R_{\sigma}=\sigma_{\rm exp}/\sigma_{\rm th}$ of cross-sections for one proton (neutron) removal from $^{24}$Si is $\sim 0.8 \pm 0.04$ ($\sim 0.39 \pm 0.04$) \cite{Gade1}. The corresponding ratio of SMEC and SM proton (neutron) spectroscopic factors is however almost constant and equals 0.987 (0.981) for $\alpha = 1$, and 0.974 (0.957) for $\alpha = 0.73$. 

Fig. \ref{fig_2} shows the $\Delta S$ dependence of the ratio of SMEC and SM g.s. spectroscopic factors for mirror nuclei: $^{28}$S and $^{28}$Mg. In these nuclei, the distribution of SM spectroscopic strength is different from that in $^{24}$Si and $^{24}$Ne. Only 44$\%$ of the proton $s_{1/2}$ spectroscopic strength is in the g.s. of $^{28}$S. On the contrary, 97.6$\%$ of the neutron $d_{5/2}$ spectroscopic strength is concentrated in the g.s. of $^{28}$S. Consequently, even in the absence of the $T=0$ component in the continuum coupling interaction ($\alpha =1$), the ratio $R_{SF}$ of the proton $s_{1/2}$ spectroscopic strengths depends on $S_p - S_n$. 
This dependence is further enhanced for $\alpha \neq 1$ (see Fig. 2b). 
One should also notice strong breaking of the mirror symmetry by comparing the ratio of proton $s_{1/2}$ spectroscopic strength 
at small $S_p$ in $^{28}$S (Fig. 2b) and the ratio of neutron $s_{1/2}$ spectroscopic strength at small $S_n$ in $^{28}$Mg (Fig. 2c). 
This effect is due to an important dependence of the $s$-wave continuum coupling on the Coulomb interaction.

In $^{28}$S, the ratio $R_{\sigma}$ of proton (neutron) removal cross-sections is $\sim 0.92 \pm 0.07$ ($\sim 0.31 \pm 0.025$) \cite{Gade1}. Again, the corresponding ratio of SMEC and SM proton (neutron) spectroscopic factors is almost constant  and equals 1.026 (1.0) for $\alpha = 1$ and 0.848 (0.975) for $\alpha = 0.73$. 

\begin{figure}[htb]
\includegraphics[width=0.83\linewidth]{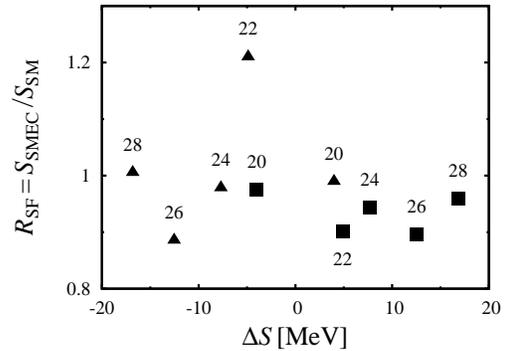}
\caption{The ratio of SMEC and SM g.s. spectroscopic factors in the chain of Ne isotopes ($20\leq A\leq 28$) is plotted as a function of the asymmetry $\Delta S$ of their experimental neutron and proton separation energies for $\alpha=0.73$. The ratios of proton (neutron) spectroscopic factors are depicted with squares (triangles). In agreement with the experimental angular momentum and parity of the nucleus $(A-1)$, the g.s. proton spectroscopic factors are $d_{5/2}$ for all considered isotopes, and neutron spectroscopic factors are  (i) $s_{1/2}$ in $^{20,26}$Ne, (ii) $d_{3/2}$ in $^{22,28}$Ne, and (iii) $d_{5/2}$ in $^{24}$Ne. For more details, see the description in the text.}
\label{fig_3}
\end{figure}

Systematics of the ratio $R_{SF}$ for even-$N$ Ne-isotopes is plotted in Fig. \ref{fig_3}. $R_{SF}$ for proton (neutron) spectroscopic factors are shown with squares (triangles) as a function of $S_p-S_n$ ($S_n-S_p$). One can see that the ratio of SMEC and SM spectroscopic factors does not exhibit any systematic tendency as a function of $\Delta S$. The fluctuations of $R_{SF}$ are small except for the neutron spectroscopic factor in $^{22}$Ne. This nucleus is however an exception in the Ne-chain because the g.s. SM spectroscopic factor is much smaller than the spectroscopic factor in the first excited state and therefore even a small spin (isospin) dependence of the proton-neutron continuum coupling and/or a small isospin-symmetry breaking term in the Hamiltonian and the continuum coupling, produce significant change of the spectroscopic factor.

In conclusion, we have consistently evaluated the fraction of single-particle spectroscopic strength which is shifted to higher excitations as a result of the coupling to the proton and neutron decay channels. We have shown that the one-nucleon spectroscopic factors in SMEC are weakly correlated with the asymmetry of neutron $S_n$ and proton $S_p$ separation energies. Strong mirror symmetry breaking, found in the ratio of SMEC and SM spectroscopic factors, appears mainly for small one-nucleon separation energies suggesting the threshold nature of this effect. Whatever the precise reasons for a strong dependence of the ratio of experimental and theoretical one-nucleon removal cross sections on the asymmetry of neutron and proton separation energies are, the explanation of this dependence does not reside in the behavior of the one-nucleon spectroscopic factors as a function of $\Delta S$. 

	This work was supported in part by the FUSTIPEN (French-U.S. Theory Institute for Physics with Exotic Nuclei) under U.S. DOE grant No.~ DE-FG02-10ER41700, by the COPIN and COPIGAL French-Polish scientific exchange programs, by the U.S. DOE contract No.~DE-AC02-05CH11231 (LBNL), and by PHC Xu GuanQi 2015 under project number 34457VA. Y.H.L. gratefully acknowledges the financial supports from the National Natural Science Foundation of China (Nos. U1232208, U1432125), Ministry of Science and Technology of China (Talented Young Scientist Program) and from the China Postdoctoral Science Foundation (2014M562481). N.A.S. acknowledges the funding of CFT (IN2P3/CNRS, France), AP th\'eorie 2014.

\bibliographystyle{apsrev4-1}

\begin{thebibliography}{} 
\bibitem{theory_tost} P.G. Hansen and J.A. Tostevin, Annu. Rev. Nucl. Part. Sci. {\bf 53}, 219 (2003);\\
J.A. Tostevin, Nucl. Phys. {\bf A682}, 320c (2001).
\bibitem{Gade1} A. Gade et al., Phys. Rev. Lett. {\bf 93}, 042501 (2004).
\bibitem{Gade2} A. Gade et al., Phys. Rev. C {\bf 77}, 044306 (2008).
\bibitem{mahaux69_b16} C. Mahaux and H.A. Weidenm\"{u}ller, {\it Shell Model Approach to Nuclear Reactions} (North-Holland, Amsterdam, 1969).
\bibitem{barz77_82} H.W. Barz, I. Rotter and J. H\"{o}hn, Nucl. Phys. {\bf A 275}, 111(1977);\\
R.J. Philpott, Fizika {\bf 9}, 109 (1977).
\bibitem{bennaceur00_44} K. Bennaceur, F. Nowacki, J. Oko{\l}owicz and M. P{\l}oszajczak, Nucl. Phys. {\bf A671}, 203 (2000).
\bibitem{Jag14} Y. Jaganathen, N. Michel and M. P{\l}oszajczak, Phys. Rev. C {\bf 89}, 034624 (2014).
\bibitem{Hagen} {\O}. Jensen, G. Hagen, M.Hjorth-Jensen, B.A. Brown, and A. Gade, Phys. Rev. Lett. {\bf 107}, 032501 (2011).
\bibitem{Flavigny} F. Flavigny, et al., Phys Rev. Lett. {\bf 110}, 122503 (2013).
\bibitem{Obertelli} C. Louchart, A. Obertelli,  A. Boudart, and F. Flavigny, Phys. Rev. {\bf 83} 011601 R (2011).
\bibitem{Oko1} J. Oko{\l}owicz, M. P{\l}oszajczak and I. Rotter, Phys. Rep. {\bf 374}, 271 (2003).
\bibitem{Wild84} B.H. Wildenthal, Progress in Particle and Nuclear Physics, ed. D.H. Wilkinson (Pergamon, Oxford, q1984) vol. II, p. 5.
\bibitem{Brown} B.A. Brown and B.H. Wildenthal, Annu. Rev. Nucl. Part. Sci. {\bf 38}, 29 (1988).
\bibitem{Ormand} W.E. Ormand and B.A. Brown, Nucl. Phys. {\bf A440}, 274 (1985).
\bibitem{OMNP12} J. Oko{\l}owicz, N. Michel, W. Nazarewicz and M. P{\l}oszajczak, Phys. Rev. C {\bf 85}, 064320 (2012).

\end{thebibliography}

\end{document}